\documentclass [12pt]{article}   
\usepackage{epsfig}
\usepackage{hangcaption}
\usepackage{array}
\newcommand{\doublespace}{
     \renewcommand{\baselinestretch}{1.5}\large\normalsize}
\newcommand{\be}{\begin{equation}}
\newcommand{\ee}{\end{equation}}
\newcommand{\bear}{\begin{eqnarray}}
\newcommand{\eear}{\end{eqnarray}}
\begin{document}
\begin{flushright}
SNUTP-99-055 
\end{flushright}

\vspace{1cm}
\begin{center}
\begin{large} 
{\bf Meson Mass at Large Baryon Chemical Potential in Dense QCD}\\
\end{large}
\vskip 1.0in
Deog Ki Hong$^a$\footnote{Email: dkhong@hyowon.cc.pusan.ac.kr},
Taekoon Lee$^b$\footnote{Email: tlee@ctp.snu.ac.kr} and
Dong-Pil Min$^{b,c}$\footnote{Email: dpmin@phya.snu.ac.kr}
\\
\vskip 1.0cm
        {\small
        {\it $^a$Department of Physics, Pusan National University,
        Pusan 609-735, Korea\\
        $^b$Center for Theoretical Physics and
        $^c$Department of Physics,\\
        Seoul National University,  Seoul 151-742, Korea}}
\end{center}
\date{}
\vskip 1cm
\begin{abstract}
We reexamine the quark mass induced term in chiral Lagrangian
in color--flavor locking phase in dense QCD,
and show that the meson mass term is determined by three independent 
invariants under chiral--axial symmetry, and a meson mass is given in
terms of the
quark mass, gap, and the chemical potential by $m_{\pi}^2\sim
m_q^2\Delta\bar{\Delta}/\mu^2\ln(\mu^2/\Delta^2)$. Thus
mesons become massless as $\mu\rightarrow\infty$.

\end{abstract}
\vskip      1cm
\pagenumbering{arabic}
\addtocounter{page}{0}

\doublespace
Cold, dense quark system with three massless quark flavors was proposed to be
in color--flavor locking  phase first  by Alford, Rajagopal and Wilczek
\cite{arw},    
and subsequently confirmed by others \cite{others}
, in which di-quark condensation
occurs in a pattern
\bear
< \chi_i^a\chi_j^b> =-< \bar{\varphi}_i^a\bar{\varphi}_j^b>=
\,\,k_1\, \delta^a_i\delta^b_j + k_2 \,\delta^a_j\delta^b_i
\label{e1}
\eear
where $\chi^a_i,\varphi^a_i$, $a$ the color index, $i$  the flavor index,
denote two-component Weyl fermions for the left-handed quarks and the
complex conjugate of
right-handed quarks, respectively.
In this phase the symmetry of dense QCD, namely, the global 
color--chiral--axial--baryon-number symmetry
$SU_c(3)\times SU_L(3)\times SU_R(3)
\times U_A(1) \times U_B(1)$ is spontaneously broken to
$SU_{c+L+R}(3)$ by the quark condensation.
The  $U_A(1)$, which  is anomalous at zero density, is a good symmetry at high
density, since instanton effects are screened out at large baryon chemical
potential $\mu$ \cite{instanton}. 
Under the symmetry quarks transform as
\bear
\chi^a_i&\rightarrow& U_{Lij} g^{ab} e^{i(\alpha+\beta)} \chi^b_j\nonumber \\ 
\varphi^a_i&\rightarrow& U^{*}_{Rij} g^{*ab} e^{i(\alpha-\beta)} \varphi^b_j
\eear
where $U_{L,R} \in SU_{L,R}(3)$, $g\in SU_c(3)$ and $e^{i\alpha}, e^{i\beta}$
for the axial and baryon number symmetry, respectively.

Upon the quark condensation $9+9$ Nambu-Goldstone bosons are generated, each 9
coming from the $\chi\chi$ and $\bar{\varphi}\bar{\varphi}$ condensation. 
In nonlinear
realization of the symmetry these Nambu-Goldstone
bosons can be represented by the  unitary matrices $U^a_i, V^a_i$ which
parameterize the coset space
$SU_c(3)\times SU_L(3)\times U_{A+B}(1)/SU_{c+L}(3)$ and $SU_c(3)\times
SU_R(3)\times U_{B-A}(1)/SU_{c+R}(3)$, respectively, and transform as
\bear
U^a_i &\rightarrow& U_{Lij}g^{ab} U^b_j e^{i(\alpha+\beta)}\nonumber\\
V^a_i &\rightarrow& U_{Rij}g^{ab} V^b_j e^{i(-\alpha+\beta)}.
\eear

Upon gauging the color symmetry gluons absorb 8 Nambu-Goldstone bosons, becoming
massive $ m_{A_{\mu}}\sim gF_{\pi}$, where $g$ is the gauge coupling,
via Higgs mechanism, and there remain 10 Nambu-Goldstone bosons (mesons) in the 
physical spectrum. Now when quarks receive small mass via the Dirac mass term
\be
{\cal L}_{m_q}= - m_{ij}\chi^a_i\varphi^a_j +h.c.
\label{dm}
\ee
these mesons, except for the one associated with the baryon
number symmetry, become massive due to the explicit breaking of the
chiral--axial symmetry. Since these mesons would be the lowest lying
excitations in cold, dense system such as neutron star their spectrum is
of considerable interest and was investigated recently by 
several  groups \cite{son-stephanov,rho,gatto}. A notable feature obtained
from the
investigation is that at large chemical potential
meson mass is independent of the gap, $m_\pi^2\sim m_q^2$, where $m_q$ denotes
quark mass, and
when there is no sextet condensation (i.e. $k_1=-k_2$)
meson mass is determined by a unique chiral--axial invariant and
all mesons become massless  as any two of the quarks go massless.

In this letter we reexamine this problem  in the framework of
chiral Lagrangian that includes the quarks as well as the Nambu-Goldstone bosons,
and reach to a conclusion  different from those obtained from the
previous investigations.
We find that meson mass is gap-dependent, $m_{\pi}^2\sim
m_q^2\Delta\bar{\Delta}/\mu^2
\ln(\mu^2/\Delta^2)$, where $\Delta$ and  $\bar{\Delta}$ denote the gaps
for particle and antiparticle respectively,
and  for arbitrary gap parameters $k_i$, 
it is  determined by three 
chiral--axial invariants,
and mesons become massless only to the leading order in
$\Delta\bar{\Delta}/\mu^2$, but massive at next leading order,
$m_{\pi}^2\sim m_q^2\Delta^2\bar{\Delta}^2/\mu^4\ln(\mu^2/\Delta^2)$, 
even if there is
no sextet condensation  and two of the quarks become massless.

We start with the chiral Lagrangian for the mesons.
The 9 chiral--axial mesons can be represented by the color singlet field
\be
 \Sigma_{ij}=U^{a}_i V^{*a}_j
 \ee
and can be described by the Lagrangian 
\bear
{\cal L}_{\Sigma} =  \frac{F_8^2}{4} {\mbox{Tr}}[ \partial_0\tilde{\Sigma}^{\dagger}
\partial_0\tilde{\Sigma} -v^2 \partial_i\tilde{\Sigma}^{\dagger}
\partial_i\tilde{\Sigma}] +  \frac{F_1^2}{2} [ (\partial_0\alpha)^2-v^2(\partial_i\alpha)^2]
+{\cal L}_m(\Sigma),
\eear
where we have kept only the leading term in derivative expansion, and
factorized the octets $\tilde{\Sigma}$ and  $U_A(1)$ part by
$\Sigma=\tilde{\Sigma} e^{2 i \alpha}$.
$F_8, F_1$ are the decay constants associated with the octets and
the singlet, respectively, and $v$ is the meson velocity which was
previously calculated to be $1/\sqrt{3}$ \cite{son-stephanov, rho}.
${\cal L}_{m}(\Sigma)$ is the
quark mass induced term, and determines the meson mass. 
Except for  ${\cal L}_{m}(\Sigma)$ the Lagrangian is invariant under
\be
\Sigma \rightarrow U_L \Sigma U_R^{\dagger} e^{2 i\theta},
\label{t1}
\ee
where $\theta$ is an arbitrary phase, 
and since the quark mass term (\ref{dm}) is also invariant under the
chiral--axial symmetry
when the Dirac mass transforms as 
\be
m \rightarrow U_L^{*} m U_{R}^{t} e^{-2 i\theta},
\label{t2}
\ee
${\cal L}_{m}(\Sigma)$ should be invariant under the combined
transformation (\ref{t1}) and (\ref{t2}). 

${\cal L}_{m}(\Sigma)$ can be expanded in powers of the quark mass matrix.
Hereafter we mean by quark mass the Dirac mass, not the Majorana mass
induced by the gap.
Unlike in zero density QCD where ${\cal L}_{m}(\Sigma)$ is linear in
the quark mass matrix, the leading term is quadratic in $m_{ij}$ due to
the absence of left-right quark condensation. It is not difficult to see that
there are three independent terms 
which are quadratic in quark mass and invariant under
 (\ref{t1}) and (\ref{t2}), and thus  ${\cal L}_{m}(\Sigma)$  is given by 
\bear
{\cal L}_m(\Sigma)&=& A \,\,[
{\mbox{Tr}}(m^t\Sigma)]^2 + B\,\, {\mbox{Tr}}[(m^t\Sigma)^2] +C\,\, 
{\mbox{Tr}}(m^t\Sigma){\mbox{Tr}}(m^*\Sigma^\dagger) + h.c.\, 
\label{form}
\eear
where $A, B$ and $C$ are constants that possibly depend on the gap and 
the chemical potential.

A convenient framework to determine these coefficients is provided by
the following chiral Lagrangian,
which is valid at energies below the scale at which the quark condensation 
occurs, for the  quarks and Nambu-Goldstone bosons before color is gauged
\bear
{\cal L}&=& i \bar{\chi}^a_i\bar{\sigma}^{\nu}\partial_\nu\chi^a_i +
\mu \bar{\chi}^a_i\bar{\sigma}^{0}\chi^a_i +  i
\bar{\varphi}^a_i\bar{\sigma}^{\nu}\partial_
\nu\varphi^a_i -
\mu \bar{\varphi}^a_i\bar{\sigma}^{0}\varphi^a_i - [m_{ij}\chi^a_i\varphi^a_j +h.c.]
\nonumber \\
&&+\left[ \chi^a_i (\Delta_\chi^\dagger)^{ab}_{ ij}\chi^b_j -
\varphi^a_i(\Delta_\varphi)^{ab}_{ ij}\varphi^b_j +h.c.\right] 
+{\cal L}_{{\mbox{{\tiny NG}}}}(U, V).
\label{l1}
\eear
Here,
\bear
 (\Delta_\chi^\dagger)^{ab}_{ij}&=&(k_1 U^{*a}_iU^{*b}_j +k_2 U^{*a}_jU^{*b}_i)
 \mbox{\bf P}_{-}(\partial) +(\bar{k}_1 U^{*a}_iU^{*b}_j +\bar{k}_2
 U^{*a}_jU^{*b}_i)\mbox{\bf P}_{+}(\partial)
 \nonumber \\
(\Delta_\varphi)^{ab}_{ ij}&=&(k_1\, V^{a}_i\,V^{b}_j \,\,+k_2\,
V^{a}_j\,V^b_i\,)
\mbox{\bf P}_{+}(\partial)\, +\,(\bar{k}_1\, V^{a}_i\,V^{b}_j \, +
\bar{k}_2\, V^{a}_j\,V^b_i\,)\,\mbox{\bf P}_{-}(\partial),
\eear
where the projection operators $\mbox{\bf P}_{\mp}(\partial)$ are given by
\be
\mbox{\bf P}_{-}(\partial)=(1+i\vec{\sigma}\cdot\hat{\partial})/2 \hspace{.5in}
\mbox{\bf P}_{+}(\partial)=(1-i\vec{\sigma}\cdot\hat{\partial})/2
\ee
and $\bar{k}_i$ denote the gap for the antiparticles \cite{ws}.
${\cal L}_{{\mbox{{\tiny NG}}}}(U, V)$, which is irrelevant for our
immediate discussion, is the usual chiral Lagrangian
for the Nambu-Goldstone bosons alone \cite{hrz,gatto1}, and also supposed to
contain  the mass term ${\cal L}_{m}(\Sigma)$. Note that $\varphi$
has negative chemical potential, as it should, since $\varphi$ 
carries negative baryon number.

In this effective Lagrangian  the meson mass term ${\cal L}_{m}(\Sigma)$
can be obtained by integrating out quarks in the constant background
of $U$ and $V$, which gives the one loop quark vacuum energy.
This may appear confusing, since the meson mass term is already present in
the Lagrangian. To avoid the confusion we must regard the effective
Lagrangian (\ref{l1}) in Wilsonian sense, namely, that 
it is obtained by integrating out
higher frequency modes to the scale where it is defined . When we integrate out
higher frequency modes further to obtain a Lagrangian defined at a lower scale,
the renormalization to the meson mass term comes in leading order
from the quark vacuum energy, and in this sense the meson mass term can be 
obtained by integrating out quarks from the UV cutoff ($\sim\mu$) to the 
scale of interest, which in this case is the gap.

Now power expansion in quark mass of the quark vacuum energy corresponds to
insertions of quark mass in the one loop bubble. Since there is no quark
propagator
connecting the left-handed quark ($\chi$) to the right-handed one
($\varphi$) at zero quark mass,
only even number of quark mass insertions are possible. This immediately
shows that the leading term in the meson mass must be quadratic in quark mass,
as already pointed out in \cite{arw}.

A straightforward calculation shows that ${\cal L}_{m}(\Sigma)$
at two mass insertions is given in the form (\ref{form}), with
the coefficients given as the sum of integrals like
\be
i \int \frac{d^{4}p}{(2\pi)^4} 
\frac{\Delta\bar{\Delta}}{[p_0^2 -\epsilon_{-}^2(p)][p_0^2
-\epsilon_{+}^2(p)]}
\label{coef1}
\ee
for $A$ and $B$, and 
\be
i \int \frac{d^{4}p}{(2\pi)^4} 
\frac{(p_0-\mu)^2 -|\vec{p}|^2}{[p_0^2 -\epsilon_{-}^2(p)][p_0^2
-\epsilon_{+}^2(p)]}
\label{coef2}
\ee
for $C$. Here
\be
\epsilon_{-}^2(p)=(|\vec{p}|-\mu)^2+\Delta^2,
\hspace{.5in}
\epsilon_{+}^2(p)=(|\vec{p}|+\mu)^2+\bar{\Delta}^2,
\ee
and
we denoted quantities proportional to the  particle gap $k_i$ by the
generic gap $\Delta$ and those for the 
antiparticle gap $\bar{k}_i$ by $\bar{\Delta}$.
An interesting feature of the integrands in (\ref{coef1}), 
(\ref{coef2}) is that
when the quark momentum in the loop is near the Fermi 
surface ($|\vec{p}|\sim
\mu$), only one quark propagator  assumes its Fermi surface value but
the other does not. 
This can be easily understood when we observe that
the quark mass term (\ref{dm}) creates a  particle and a 
Dirac hole out of the perturbative vacuum (which is the state in which 
free massless quarks are filled up to the Fermi surface), 
and thus when the particle is near the
Fermi surface, the Dirac hole is $2\mu$ off from it.

Performing the integration, the details of which will be given elsewhere
\cite{hlm},
we obtain
\be
A\sim B\sim \Delta\bar{\Delta}\ln(\mu^2/\Delta^2),
\hspace{.3in} C\sim (\Delta^2\bar{\Delta}^2/\mu^2)\ln(\mu^2/\Delta^2).
\label{e13}
\ee

It is interesting to compare this result with the previous calculations.
If we assume no sextet condensation, that is, $k_1=-k_2$, we get
$A=-B +O(\Delta^2\bar{\Delta}^2/\mu^2)$, and thus
\be
{\cal L}_m(\Sigma) \sim \Delta\bar{\Delta}\ln(\mu^2/\Delta^2)\left[
{\mbox{det}}(m\Sigma) {\mbox{Tr}}[
m^{-1} \Sigma^{*}] +h.c.\right] + O[(\Delta^2\bar{\Delta}^2/\mu^2)
\ln(\mu^2/\Delta^2)].
\ee
The leading term in $\Delta\bar{\Delta}/\mu^2$, 
which vanishes if any two of the
quarks become massless,
is what was found in  \cite{son-stephanov},
except for the prefactor,
which in their calculation was
found to be $\sim \mu^2$ instead of our $\sim 
\Delta\bar{\Delta}\ln(\mu^2/\Delta^2)$
(To obtain the form in the reference, one must replace $m \rightarrow
m^{*}$ and do a field redefinition ${\mbox{det}}(\Sigma^{*}) \Sigma
\rightarrow\Sigma$ to compensate the differences in definition).
This also shows that, contrary to the claim in  \cite{son-stephanov},
the leading term is not the unique one allowed by the chiral--axial symmetry,
and even if two of the quarks become massless the mesons can be massive at
next leading order in $\Delta\bar{\Delta}/\mu^2$.
We also notice that the meson mass term discussed in \cite{gatto},
which is 
in non-analytic form in the quark mass, does not appear. From the discussion
so far,
it is clear that such term cannot be present in the effective Lagrangian.

Eq. (\ref{e13}) also implies an interesting behavior for the meson mass at
large chemical potential. From  (\ref{form}) we see $m_\pi^2 \sim
m_q^2\Delta\bar{\Delta}\ln(\mu^2/\Delta^2)/
F_\pi^2$, where $F_\pi$ is the pion
decay constant, which can  be 
calculated also within the effective Lagrangian (\ref{l1}).
The matrix element in the definition of the decay constant 
associated with a current
\be
<0| J_\mu(0)|\pi(p)> = i F_\pi p_\mu
\ee
can be evaluated in one loop using the quark--quark--Nambu-Goldstone boson 
vertices given in (\ref{l1}). Performing the loop calculation we find 
$F_\pi \sim \mu$, in agreement with the previous calculations
\cite{son-stephanov,rho}. This then gives
\be 
m_\pi^2 \sim
m_q^2\Delta\bar{\Delta}\ln(\mu^2/\Delta^2)/\mu^2
\ee
which shows that  the mesons become massless at asymptotically large chemical
potential.

\vskip .5in
\noindent
{\bf Acknowledgments:} 
This work was supported in part by the Korea Science and  Engineering
Foundation (KOSEF).


\begin{thebibliography}{10}
\bibitem{arw} M. Alford, K. Rajagopal, and F. Wilczek, 
Nucl. Phys. B {\bf 537} (1999) 443.
\bibitem{others}
D.~K. Hong,  {\tt hep-ph/9905523};
T. Sch\"afer, {\tt hep-ph/9909574};
N. Evans, J. Hormuzdiar, S.~D.~H. Hsu and M. Schwetz,
{\tt hep-ph/9910313.}
\bibitem{instanton}
R.~D. Pisarski and D.~H. Rischke, Phys. Rev. Lett. {\bf 83}
(1999) 37;
R. Rapp, T. Sch\"afer, E.V. Shuryak and M. Velkovsky,
{\tt hep-ph/9904353}.
\bibitem{son-stephanov} D.T. Son and M.A. Stephanov, {\tt hep-ph/9910491}.
\bibitem{rho} M. Rho, A. Wirzba, and I. Zahed, {\tt hep-ph/9910550}:
          After submission of our paper, a revised version of this reference
	  has appeared in which it was claimed that mesons remain massless
	 to order $m_q^2$.
\bibitem{gatto} R. Casalbuoni and R. Gatto, {\tt hep-ph/9911223}.
\bibitem{ws} T. Sch\"afer and F. Wilczek, Phys. Rev. D {\bf 60} (1999) 114033.
\bibitem{hrz}D.~K. Hong, M. Rho, and I. Zahed,
Phys. Lett. B {\bf 468} (1999) 261, {\tt hep-ph/9906551}.
\bibitem{gatto1} R. Casalbuoni and R. Gatto, Phys. Lett. B {\bf 464}
(1999) 111, {\tt hep-ph/9908227}.
\bibitem{hlm} D.~K. Hong, T. Lee and D.-P. Min,  in preparation.

\end{thebibliography}

\end{document}